\pdfoutput=1 
\documentclass[prodmode,permissions]{acmsmall-ec13}

\usepackage[ruled]{algorithm2e}

\usepackage{enumitem}
\usepackage{indentfirst}
\usepackage{graphicx}
\renewcommand{\algorithmcfname}{ALGORITHM}
\usepackage{amsmath,amsfonts,amssymb,bbm} 
\usepackage[numbers,sort&compress]{natbib} 
\usepackage{fancyhdr}
\usepackage{lastpage}
\SetArgSty{textrm}  
\SetAlFnt{\small}
\SetAlCapFnt{\small}
\SetAlCapNameFnt{\small}
\SetAlCapHSkip{0pt}
\IncMargin{-\parindent}
\usepackage{multicol}
\usepackage{xcolor}

\begin{document}

\title{The Game Performance Index for Mobile Phones} 

\author{Hesham Dar, James Kwan, Yang Liu, Omiros Pantazis, Robert Sharp \affil{Samsung Research UK}}

\begin{abstract}
With the recent increase in the quantity of high fidelity games appearing on mobile devices and the recent trend of gaming focused mobile devices, there is a new requirement for a clear and comprehensive measure of the quality of gaming performance on the mobile device platform. This paper proposes a conceptual framework for a user-experience and user-perception based set of performance measures for mobile devices. This paper presents a specific implementation and measurement use case which has been beneficial to Samsung Electronics when applied to our own product range, allowing us to better understand and quantify device performance. We believe that the methods outlined are potentially useful to the consumer, by providing an understandable public facing score for device performance to guide consumers with purchasing decisions. The methods may be useful to game developers and could better enable the developer to add new richer game features based on the performance of the device.
\end{abstract}

\maketitle

\begin{multicols}{2}
\section{Introduction}

As the capability and processing power of mobile devices have increased, so too have the complexity and fidelity of games that are available on these devices. This has been particularly notable in recent years where a tipping point has been reached; games originally designed for the traditional platforms such as game consoles and PC platforms are now appearing on mobile devices, trying to offer to the user a similar gaming experience. This transition poses a significant challenge to device manufacturers as there is no shortage of resource demanding, high fidelity games available on game consoles and PC platforms that could potentially be migrated to mobile as soon as the devices reach a minimum accepted threshold of processing power. These games are initially developed on console/PC platforms with very different constraints or thermal control, power consumption and form factor. Moving a console game to a mobile device typically requires extensive game optimisation. Despite this optimisation, games are still prone to performance related issues on the mobile device and may struggle to reach the same high quality of user-game experience when compared with other more powerful platforms. \par
Game performance measurement is of critical importance to many different stakeholders in the gaming-space. Device manufacturers and game developers both strive to provide the best gaming experience possible and may benefit from a measurement which reflects that experience as closely as possible. Consumers have a huge amount of choice when it comes to selecting mobile game devices. An easy to understand measure may help considerably to inform purchasing decisions where the consumers are able to judge the performance of a device when playing their game of choice. \par

Approaches to measuring game performance and the gaming performance capability of devices are well established for traditional platforms, and a variety of approaches are employed. Current measurements tend to focus on the processing power of the devices with the assumption that this measure correlates well with the user-game experience. Mobile device specific factors (such as power consumption) are not typically reported. These traditional game performance measurement approaches have, in many cases, been used with little or no modification for mobile devices. Some of these measurements are not well suited for mobile phone device (i.e. peak performance measurements rather than sustained performance) and the quality of the user-game experience may be impacted by other factors of the device which are not measured and do not affect other platforms (e.g. battery life). \par
This paper proposes a framework for defining performance of mobile games which attempts to better reflect the actual user-game experience. This is a generalised approach based on a review of the domain, internal practices within Samsung Electronics in addition to key insights derived from the gaming community itself.

\section{Background}

Game performance measurement exists in a number of different forms and the approach is usually specific to a given use case. The most common performance measurement provided to the consumer is a device comparison presented in technical articles, forums and blogs. This is generally the shallowest level of game performance measurement, provided as subjective statements describing the feelings of reviewer while using the device. The review is typically accompanied with quoted device specifications including details of the processor used, memory etc., and in the case of mobile devices the battery capacity. While this approach does provide some valuable information to the consumer, the reviews do not provide clear information for the purpose of comparison; lacking both quantitative information and clear testing procedure. \par
Game performance measurement that uses a testing procedure is typically referred to as benchmarking. Benchmarking falls into one of two categories: synthetic benchmarks and application or "real game" benchmarks. 

\subsection{Synthetic Benchmarks}

Synthetic benchmarks are general tests of a system's performance. The synthetic benchmark is typically a custom made simulation that stresses a particular function of the system as it is executed and a measurement is made to grade the performance of the system. \par
For game performance benchmarks, a benchmarking program imposes a load on the gaming device which attempts to reproduce the processing loads of a game in a realistic manner. While processing the simulated load, different performance characteristics of the device are measured and combined to produce a score from a number of sub-tests. Each sub-test is measuring a specific component during the simulation such as the CPU. Popular mobile synthetic benchmarks include 3DMark and AnTuTU (for both the PC and mobile platforms). Synthetic benchmarks have a number of advantages, notably having the benefit of being able to run simulations of a variety of loads, test specific components of the system under a range of settings and are repeatable. \par
There are many benefits to synthetic benchmarks for gaming although they do suffer from a number of issues, particularly for a mobile gaming use case. Simulated loads are designed to be realistic but it is difficult for them to account for every type of load that would occur in the natural gameplay of a particular game. Many of the leading synthetic benchmark solutions present an unbounded score made up of a number of constituents that can be difficult to interpret and the results are not necessarily indicative of the user-game experience. It is not surprising that these types of results do not generally reflect the gameplay experience as these benchmarks are primarily aimed at the device or component manufacturer. These tests tend to exercise peak loads or peak throughput and on a mobile device do not represent real world scenarios because of the power and thermal constraints where sustained performance becomes more important.

\subsection{Application Benchmarks}

Application benchmarking for game performance avoids the problem of having to develop a proxy for a game by testing with the game itself. Performance measures are recorded at the game level and as such are easier to interpret. Some games include bespoke benchmarking modes, however for the majority of games the test procedure requires actual user-game play. A game session is played on a device in a "natural" manner and game level performance metrics are collected. These measures heavily focus on frame rate, and in many cases an average frame rate is reported, while being indicative of the game performance, they are not comprehensive. \par
Due to the way that application benchmarking has evolved, practices that are applied to mobile game benchmarking align heavily with the PC and home console platforms, measurements do not account for the additional concerns mobile gamers have in terms of power consumption and device heat. \par
There are a number of approaches and solutions available for measuring game performance of mobile devices, while there is no accepted standard, it is often easy to omit certain aspects to simplify comparisons. The mobile gaming space represents a new challenge for the existing PC and console gaming benchmarking. The way in which a user interacts with a mobile device, when compared to traditional gaming platforms, requires the consideration of new mobile gaming specific metrics.

\section{Game Performance Index}

There has been much research on benchmarking in a variety of domains \cite{dai2019benchmarking} and certain characteristics of this research may be desirable for mobile gaming performance measurement. We have adopted similar principles when designing our game performance measurement index:

\begin{enumerate}[label={\arabic*.}]
	\item {\bf Relevance}: The index should measure all features that are relevant to the user-game experience.
	\item {\bf Representativeness}: Index performance metrics should be broadly accepted by industry and academia.
	\item {\bf Equity}: All devices should be fairly compared.
	\item {\bf Repeatability}: Index results can be reproduced and verified by multiple parties.
	\item {\bf Practicality}: Index tests are straightforward to run and the data reasonable to collect.
	\item {\bf Transparency}: Index metrics should be easy to understand.
\end{enumerate} 

While this paper does not explicitly cover each of these characteristics as the framework described is conceptual, as a whole it has been designed with each in mind. \par
Leveraging insight into the gaming domain from passionate users, internal engineering teams, in addition to insights from mobile gaming data, we have constructed this framework for measuring mobile gaming performance in a generalised way such that it is useful and can also be adapted for specific uses if necessary.

\subsection{Performance Categories} 

During analysis of this problem we considered the user perspective, and grouped performance related measures into categories that reflect distinct components of the game experience. This was performed by answering the following questions:

\begin{enumerate} [label={\arabic*.}]
 	\item Is this particular factor relevant to the experience of a game?
	\item Can the "performance" of the factor vary in some way?
\end{enumerate}

By relevancy we consider whether the enhancement or detriment of a factor can lead to increased or reduced enjoyment of the gaming experience. This may be both direct and indirect, for example, the smoothness of the game's frame rate directly affects the user-game experience \cite{claypool2006effects}, whereas the battery drainage indirectly affects the experience.
The game performance measurement for mobile devices has been decomposed into performance categories which are termed Main Indices, as follows:

\begin{itemize} [label=$\bullet$]
 	\item Visual smoothness
	\item Graphical quality	
	\item Battery
	\item Temperature
	\item Swiftness
	\item Responsiveness
\end{itemize}

These groups, while not fully independent of one another functionally, are distinct in terms of how they are perceived. For example, a user's perception of the battery efficiency is separate from how detailed the quality of textures is. 
In the following sections we provide a brief description of each factor, outlining how it relates to the user-game experience, and in some instances, what measurements may be taken to capture the performance. Multiple game performance measures may correspond to each category, and a strategy for combining measures is also required.

\subsubsection{Visual Smoothness} 
Frame rate based measures are most often used for general measures of "game performance". Visual smoothness is a category designed to encompass all aspects that relate to the smoothness of animations such as the magnitude and consistency of the frame rate. Poor visual smoothness is detrimental to the user-game experience in a number of ways and is typically described with language such as "choppy" and "sluggish". When the frame rate is significantly low, the illusion of fluid motion in animations breaks and individual frames can be discerned by the user. Low frame rates and dropped frames occuring unpredictably, especially in graphically intensive scenes, can lead to players taking incorrect actions, which is particularly important for many real-time competitive games. The standard performance measures used for games include average frame-per-second (FPS), FPS percentiles, and FPS stability.

\subsubsection{Graphical Quality}
The look of a game can be as important as the feel of it, where engaging gameplay is complemented by compelling visuals to bring a satisfying experience to the users. With many games pushing the limits on how detailed or realistic their environments are, having a good sense of this quality is needed when attempting to capture the overall quality of the gaming experience. The purpose of this index category is to amalgamate a number of measures of graphical quality. Measures in this category could include: dynamic range, texture quality, antialiasing, and effects quality. \par
Within this category there are other graphical factors external to the game itself which relate to the display technology and must be considered. The size and pixel density are properties of a device that may affect the user-game experience; with larger, high pixel density displays able to show detailed visuals clearly.
 
\subsubsection{Battery}
The Battery contributes an important part to the mobile gaming experience despite being external to the game itself. This is one of the factors which differentiate the mobile gaming experience from other gaming devices such as consoles (when the device is not externally powered).\par
The battery level of the phone is a hard constraint on the duration of a gaming session and can be the cause of a session finishing early either due to concerns that the level is too low, or the device running out of power completely, both of which are frustrating experiences to users. The battery life is a key game performance indicator which should be carefully measured and monitored. 

\subsubsection{Temperature}
One of the main performance limiters for mobile gaming is temperature. Unlike other gaming platforms such as PC, both the CPU and GPU are often running below maximum clock frequencies due to temperature constraints. The form factor of mobile devices means that users are directly affected by high temperatures, potentially making  the user feel uncomfortable \cite{KangSamsungThermal19} and stop playing the game earlier than intended. In a comprehensive measure of performance for mobile gaming, temperature is therefore an important factor that should be accounted for.

\begin{figure*}[t]
\centering
  	 \includegraphics[width=\textwidth]{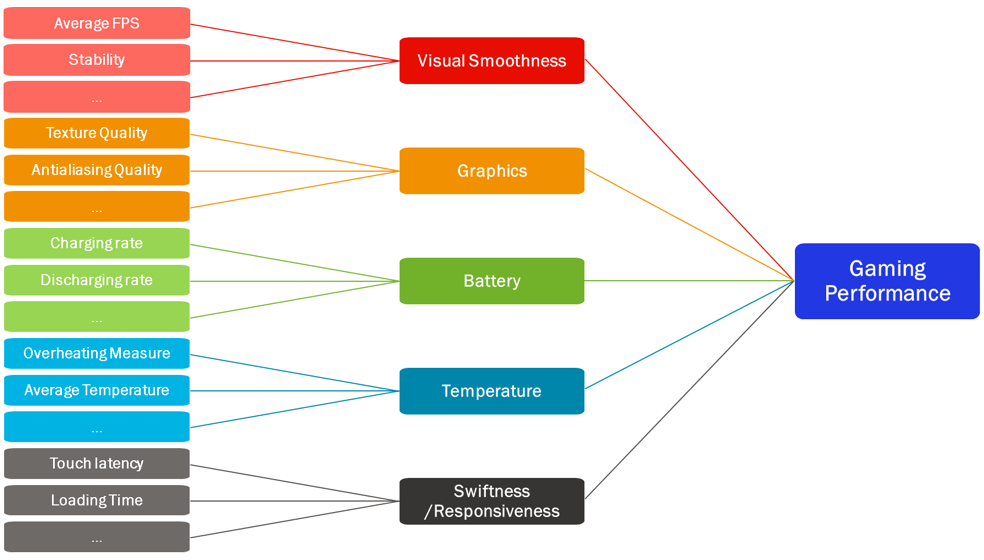}
  	  \caption{Game performance index scoring framework showing example sub-metrics being collected for each main index category to produce the overall score.}
  	  \label{fig:gpi_framework}
\end{figure*}

\subsubsection{Swiftness}
Swiftness encompasses measures that relate to performance factors that result in the user being made to wait. A prime example of this is the initial launch time of a game. Device performance is a key factor, a better performing device would be able to load a game faster, which in turn allows the user to enter the gaming session earlier, leading to an improved user-experience. While this may be less of a concern on other platforms such as PC and console, gaming on mobile has pushed the concept of "gaming on the go" where many users play games in short bursts. In this scenario, long waiting times before being able to engage with the game can be particularly frustrating to users. \par
This category includes all measures relating to waiting time, such as the initial loading time and scene loading within a gaming session. 

\begin{figure*}[t]
\centering
  	 \includegraphics[width=\textwidth]{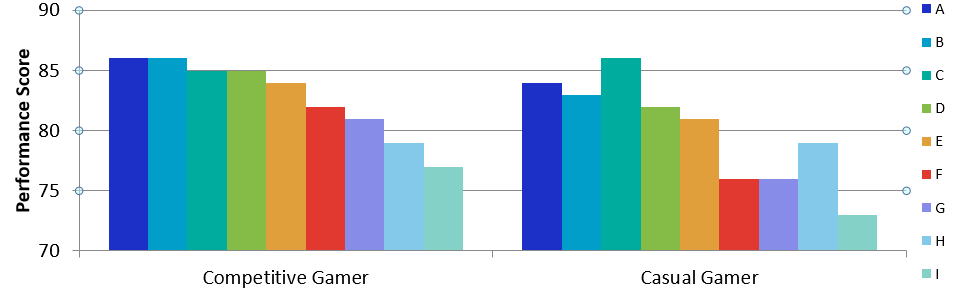}
  	  \caption{Overall game performance index scores for 9 recent flagship devices scored under a competitive and casual gamer perspective.}
  	  \label{fig:gpi_example}
\end{figure*}

\subsubsection{Responsiveness}
Device control is as important to the user-game experience as the quality and smoothness of the visuals. A good quality game with a poor control scheme, or unresponsive controls can leave the player feeling detached and make the overall experience less enjoyable \cite{eg2018playing}. A key factor of the control experience is the latency, or more specific to mobile phones, the screen touch latency. This is the fraction of time that a device needs to respond to a physical event triggered by the user. There has been much interest in the area of human perception of touch latency, with some studies showing reliable perception as low as 5-10ms \cite{ng2012designing}. This has become more important in recent years with the introduction of more competitive, "fast twitch" games that have traditionally been popular on PC and consoles, where any noticeable latency between a user's input and that input being registered can cause the user to lose a game play.

\subsection{User-Game Experience Mapping Logic}

Having defined the measurement categories that are central to the user-game experience we further align the game performance index by incorporating information from user surveys, studies, and expert consensus. This is performed at multiple levels:

\begin{itemize} [label=$\bullet$]
 	\item {\bf Metric Value to Sub-Index score mapping}: A mapping function that takes as input the raw metric value (e.g. Average FPS) and produces a Sub-Index score. 
	\item {\bf Sub-Index scores to Main Index score mapping}: A mapping function which takes as input the scores of each associated metric (e.g. Visual Smoothness from Average FPS and FPS stability) and produces a Main Index score.
	\item {\bf Main Index score to Game Performance Index mapping}: A mapping function which takes as input the Main Index scores and produces an overall Game Performance Index score.
\end{itemize} 
The user-experience mapping also has the benefit of normalising recorded metrics such that they are comparable to one another. A visual example of these relations is shown in Figure \ref{fig:gpi_framework}.

\section{Use Case}
The game performance measurement framework has been described conceptually in the previous section. A cursory implementation of the framework has been used to evaluate device performance on recent (2018 onwards) flagship mobile phones. The goal was to better understand the differentiating features of our mobile device product line with respect to gaming performance while considering the different types of end-user. \par
We collected test session data by recording performance measures from natural gameplay sessions for a number of popular high fidelity games. This was performed in a controlled manner to ensure the data was as representative as possible by controlling factors such as general device settings, in-game settings, session length, etc. \par
Metric value to Sub-Index mapping was handled on a case by case basis, using insight from engineering teams, subject matter experts or from data (where appropriate) to provide bounds for each measure. User-experience mapping logic was based on two key demographics (gamer types) that we wished to consider "Competitive" and "Casual" gamers. These were characterised by consensus among experts and appropriate main index weightings designed accordingly. The characterisations were given as broad descriptors:

\begin{itemize} [label=$\bullet$]
 	\item Competitive gamers, typically playing fast paced, real time shooting or massively multiplayer online games would value very smooth visuals with moderate graphics and have little concern for battery drainage.
	\item Casual gamers, typically playing slow paced puzzle games would value good battery life, moderate smoothness and moderate graphics.	
\end{itemize}

Multiple gaming sessions were evaluated under these criteria and median scores given to account for the natural deviation in performance between natural gaming sessions. \par
In Figure \ref{fig:gpi_example} we present the results of this investigation with overall game performance scores for an anonymised set of devices. From this we see relative changes in performance between the competitive and casual categories. Notably under a competitive setting devices A and B performed best with a score of 86 each, conversely, when measured under a casual setting device C performed best with a score of 86. \par

All devices in the sample set were flagship mobile phones and the performance across the board was relatively high. The granularity of the approach allowed for the majority of devices to be distinguishable in terms of performance. This has been valuable in providing actionable insight to related groups such as engineering teams optimising devices for gaming, and marketing who wish to promote the gaming capability of certain devices. \par

\section{Conclusion}

A new approach to measuring game performance for mobile devices is presented which offers two distinct benefits over the traditional approaches to game performance measurement:

\begin{enumerate} [label={\arabic*.}]
 	\item Comprehensive to the mobile gaming experience: Scores acknowledge and encompass the factors that are critical to mobile gaming but often not considered as strongly such as battery and temperature.
	\item User-experience driven: Scores capture what is important to the user through sensibly designed performance categories and direct mapping at multiple stages in the framework.
\end{enumerate}

We have shown how an implementation of such an approach has been useful within Samsung for helping to understand the product differentiation of mobile devices, accounting for different types of gamers.\par
We also consider this approach to be beneficial to other groups, such as game developers who can measure the user-game experience their games offer across multiple devices in a structured manner. Game developers may also use the measurement to identify the types of devices that best run their games. The approach should allow other groups to tailor measures to their own specific needs.

\bibliographystyle{ACM-Reference-Format-Journals}
\bibliography{sgpi-bibfile}

\end{multicols}
\end{document}